\documentclass[twocolumn,showpacs,amsmath,amssymb,prb]{revtex4}

\usepackage{graphicx}
\usepackage{bm}

\begin{document}

\title{Nonequilibrium charge transport in quantum SINIS structures}
\author{N. B. Kopnin $^{(1,2)}$}
\author{J. Voutilainen$^{(1)}$}
\affiliation{$^{(1)}$ Low Temperature Laboratory, Helsinki
University of
Technology, P.O. Box 2200, FIN-02015 HUT, Finland,\\
$^{(2)}$ L. D. Landau Institute for Theoretical Physics, 117940
Moscow, Russia}

\date{\today}

\begin{abstract}
Charge transport in a high-transmission single-mode long SINIS
junction (S stands for superconductor, I is an insulator, and N is
a normal metal) is considered in the limit of low bias voltages
and low temperatures. The kinetic equation for the quasiparticle
distribution on the Andreev levels is derived taking into account
both inelastic relaxation and voltage-driven Zener transitions
between the levels. We show that, for a long junction when the
number of levels is large, the Zener transitions enhance the
action of each other and lead to a drastic increase of the dc
current far above the critical Josephson current of the junction.
\end{abstract}

\pacs{73.23.-b, 74.45.+c, 74.78.Na}

\maketitle

\section{Introduction}

Weak links consisting of superconducting (S) and normal metal (N)
parts separated by insulators (I) are the subject of intensive
experimental and theoretical studies. Transport through these
systems is determined by several factors, such as transparency of
the contacts, specifics of the weak link area and energy
relaxation in the junction, leading to a series of nonlinear
characteristics in the current-voltage relation. In stationary
regime, numerous configurations have been analyzed (see
Ref.~[\onlinecite{Kupriyanov}] for review), revealing the
importance of both Andreev \cite{Andreev} and normal reflection
processes taking place in the contact.

When the injection rate of new particles into the junction is
greater than the corresponding rate of inelastic relaxation the
nonequilibrium effects are to be taken into account. In realistic
junctions with high current density and especially at low
temperatures the inelastic relaxation may become less effective,
resulting in a strong nonequilibrium which crucially affects the
current transported through the contact. For even a small bias
voltage below the energy gap, the oscillating Josephson
supercurrent may be accompanied with a nonzero dc component
corresponding to the dissipative processes. Nonequilibrium
situations in various SINIS-type junctions ranging from a point
contact to a ballistic junction with finite length have been
studied by many authors, the discussion to some extent has also
concerned\cite{GunsZaik94,AverBard95,AverBard96} the inelastic
relaxation effects. As is well known, the dc component exhibits a
subgap structure at bias voltages $eV=|\Delta|/n$ which is
associated with multiple Andreev reflections
(MAR)\cite{BTK,Octavio83,GunsZaik94}. The quasiparticles trapped
in the junction are accelerated by the applied voltage, while, for
each cycle of repeated electron-hole reflections at the two NS
interfaces, the energy of the particle increases by $2eV$ until
the accumulated energy enables it to escape the pair-potential
well. This works in a broad voltage range, but becomes more and
more complicated when relaxation effects are included or the
transparency at the interfaces differs from unity. The low-voltage
MAR process for $eV\ll \Delta$ in a ballistic contact is
equivalent to the spectral flow along the Andreev energy levels
\cite{KV03} where the phase difference $\phi$ adiabatically
depends on time $\phi =2eV/\hbar +\phi_0$. However, for a
non-ideal transparency, the energy levels are separated from each
other by minigaps (see the next section) which suppress the
transitions from one level to the next thus cutting the spectral
flow off. As a result, for very low voltages, the dc current is
small for contacts with any realistic transparency ${\cal T} \ne
1$.

The interlevel transitions can take place by means of Zener
tunneling near the avoided crossings of the Andreev levels; they
restore the spectral flow and give rise to a finite dissipative dc
current. The Zener processes are more simple in short junctions
(point contacts) where only two Andreev states corresponding to
particles travelling in opposite directions exist; these levels
have only one minigap at the phase difference $\phi =\pi$. Effects
of Zener tunneling on the transport properties of quantum point
contacts have been studied in
Refs.~[\onlinecite{AverBard95,AverBard96}]; the dc current was
found to have an exponential dependence on voltage in the
low-voltage limit.

For junctions where the center island has a length $d$ longer than
the superconducting coherence length $\xi$ the number of levels is
proportional to the ratio $d/\xi$ and can, thus, become large. If
the transparency is not exactly unity, these levels are separated
by minigaps at $\phi = \pi k$, where $k$ is an integer. In
practice, such SINIS structures can be made of a carbon nanotube
or semiconductor nanowire placed between two superconductors as in
Refs.~[\onlinecite{nanotubes,nanowire}]. In
Ref.~[\onlinecite{KMV06}] this type of junctions was suggested as
a realization of a quantum charge pump where the minigaps were
manipulated by the gate voltage being sequentially closed in
resonance with the Josephson frequency. In the present paper we
consider the low-temperature charge transport in these junctions
for constant bias and gate voltages. We derive the effective
kinetic equation for the quasiparticle distribution on the Andreev
levels taking into account both the inelastic relaxation on each
level and the Zener transitions between the neighboring levels and
demonstrate that the voltage-driven Zener transitions from one
level to the next enhance the action of each other and lead to a
drastic increase of the dc current as the transition probability
grows with the applied voltage.

We begin with a brief description of the spectral properties of
double-barrier SINIS structures in Section \ref{sec-model}. In
Section \ref{sec-kinetic} we derive the kinetic equation that
determines the distribution function on the Andreev levels in the
presence of Zener transitions and inelastic relaxation. In
Sections \ref{sec-current} and \ref{sec-discuss} we calculate the
dc current and discuss the results.

\section{Model}\label{sec-model}

We consider a quantum SINIS contact consisting of two
superconducting leads connected by a normal conductor that has a
single conducting mode. The insulating barriers have a high
transparency such that the contact is nearly ballistic. In this
Section we briefly summarize the spectral properties of SINIS
contacts which are important for the transport characteristics. It
is well known that the supercurrent  flowing through such contact
is determined by the Andreev states formed in the normal conductor
and extended into the superconducting leads. The states can be
described by the Bogoliubov--deGennes equations
\begin{equation}
\left[ -\frac{\hbar ^2}{2m}\frac{d^2}{dx^2}-E_{F} +U(x)\right]\hat
\sigma _z \hat\psi +\hat H \hat \psi  =\epsilon \hat \psi \ ,
\label{eqBog}
\end{equation}
where $\hat \sigma_z$ is the Pauli matrix in Nambu space, and
\[
\hat \psi =\left(\begin{array}{c} u\\ v\end{array}\right)\, ,\;
\hat H=\left(\begin{array}{cc} 0&\Delta \\
\Delta ^* &0 \end{array}\right) \ .
\]
The superconducting gap is $\Delta =|\Delta|e^{\pm i\phi /2}$ for
$x>d/2$ and $x<-d/2$, respectively, while $\Delta =0$ for
$-d/2<x<d/2$. For simplicity we model the normal reflections at
the interfaces as being produced by $\delta$-function barriers $
U(x)=I\delta (x-d/2)+I\delta (x+d/2)$ assuming that the
quasiparticle velocity in the superconducting leads is the same as
in the normal conductor.


In the normal region the particle, $e^{\pm iq_+ x}$, and hole,
$e^{\mp iq_- x}$, waves have amplitudes $u^\pm$ and $v^\mp$,
respectively. The upper or lower signs refer to the waves
propagating to the right $\hat \psi ^>=\left(u^+, v^-\right)$ or
to the left $\hat \psi ^< =\left(u^-, v^+\right)$. The particle
(hole) momentum is $q_\pm =k_x\pm \epsilon/\hbar v_x$ where $v_x$
is the quasiparticle velocity of the mode and $k_x=mv_x/\hbar $.
Scattering at the right and left barriers couples the amplitudes
of incident and reflected waves \cite{Beenakker}:
\begin{equation}
\hat \psi ^< _R=\hat S^R\hat \psi ^>_R , \; \hat \psi ^> _L=\hat
S^L\hat \psi ^<_L ; \;
\hat S=\! \left( \begin{array}{cc} S_{N}e^{i\delta} & S_{A}e^{i\chi} \\
S_{A}e^{-i\chi} & S_{N}e^{-i\delta}\end{array}\right) .
\label{S-R,L}
\end{equation}
The scattering matrices for the right and left barriers are $\hat
S^R=\hat S(\chi _R)$ and $\hat S^L=\hat S(\chi _L)$, respectively,
where $\chi _L=-\phi /2$ while $\chi _R=\phi /2$. The scattering
matrices are unitary $\hat S^\dagger \hat S =1$ because of
conservation of the quasiparticle flux. Components of the $\hat S$
matrix for $\delta$-like barriers and energies
$|\epsilon|<|\Delta|$ are \cite{BTK}
\[
S_{N}=-\frac{(U^2-V^2)|Z|\sqrt{Z^2+1}}{U^2+(U^2-V^2)Z^2} ,\; S_{A}
=\frac{UV}{U^2+(U^2-V^2)Z^2} .
\]
Here $Z=mI/\hbar ^2 k_x$ is the barrier strength and $
U=2^{-1/2}[1+i\sqrt{|\Delta|^2-\epsilon ^2}/\epsilon ]^{1/2}$,
$V=U^*$. The scattering phase $\delta$ is introduced through $
\cot \delta =Z $. Applying the scattering conditions at both ends
of the normal region one can derive a compact equation for the
spectrum of a SINIS contact \cite{KMV06}
\begin{equation}
|S_{N}|^2\sin ^2 \alpha ^\prime +|S_{A}|^2\cos ^2(\phi /2)
=\sin^2(\beta +\gamma) \ . \label{determinant2}
\end{equation}
Here $ \alpha =k_xd$, $\alpha ^\prime =\alpha +\delta $, and
$\beta = \epsilon d/\hbar v_x $. The phase $\gamma$ is defined as
$S_{N}= e^{i\gamma}|S_{N}|$. For short contacts $d\ll \hbar
v_x/|\Delta |$, the spectrum has two branches varying from
$\epsilon = \pm |\Delta|$ at $\phi =0$ and separated from each
other by minigaps at $\phi = \pi +2\pi k$. The energy spectra of
SINIS contacts in various limits have been extensively studied by
many authors \cite{spectrum}.

\begin{figure}[tbh]
\centerline{\includegraphics[width=0.6\linewidth]{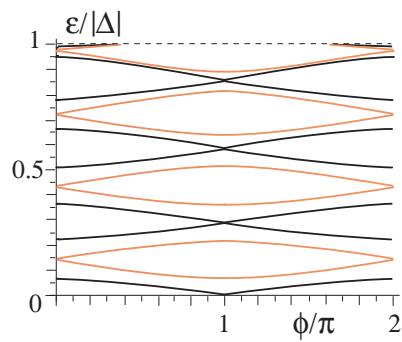}}
\caption{(Color online) Examples of the spectra, Eq.\
(\protect\ref{determinant2}), for a long SINIS contact with
$Z=0.5$ and $|\Delta| d/\hbar v =10$. Dark (black online) lines:
resonance $\sin \alpha^\prime =0$, the gaps disappear for $\phi =
\pi +2\pi k$; light (red online) lines: anti-resonance $\cos
\alpha^\prime =0$, the gaps disappear for $\phi =2\pi k$.}
\label{fig-spectrum}
\end{figure}

In what follows we focus on long contacts, $d\gg \hbar v_x/|\Delta
|$ which have a large number of levels $N\sim d|\Delta |/\hbar
v_x$. These levels split off from the states with $\epsilon =\pm
|\Delta|$ and fill the energy interval $-|\Delta| <\epsilon
<|\Delta|$ with spacings of the order of $\hbar v_x/d$. Examples
of the spectra are shown in Fig.~\ref{fig-spectrum}. Each level is
a function of the phase difference $\phi$; its range of variation
is of the order of the interlevel spacing. The levels approach
each other more closely at $\phi =\pi k$ where they are separated
by minigaps. All minigaps at $\phi =\pi (1+2 k)$ disappear for the
resonance condition, $\sin \alpha ^\prime =0$. Similarly, all
minigaps at $\phi = 2\pi k$ disappear for anti-resonance, $|\sin
\alpha ^\prime |=1$. This follows from Eq.\ (\ref{determinant2})
due to unitarity $|S_{N}|^2+|S_{A}|^2=1$. The low-energy levels,
$\epsilon_l \ll |\Delta|$, $\gamma \ll 1$ have the form
\begin{equation}
\epsilon_l =\pm \epsilon _0 +\pi \hbar v_x l/d \label{spectrum1}
\end{equation}
where
\begin{equation}
\epsilon_0 =\frac{\hbar v_x}{d} \arcsin \sqrt{{\cal T}^2\cos
^2\frac{\phi}{2}+(1-{\cal T}^2)\sin ^2\alpha ^\prime} \ ,
\label{spectrum-sym}
\end{equation}
$l$ is an integer, $ {\cal T}=(1+2Z^2)^{-1} $ is the transmission
coefficient of the contact. The energy gaps at $\phi =\pi (1+2 k)$
are all equal,
\[
\delta \epsilon_{\pi}=(2\hbar v_x/d)\arcsin \left[ |\sin \alpha
^\prime |\sqrt{1-{\cal T}^2}\right] \ .
\]
The gaps $\delta \epsilon_{2\pi}$ at $\phi =2\pi k$ are given by
the same expression where $|\sin \alpha ^\prime |$ is replaced
with $|\cos \alpha ^\prime |$. For a transparent contact, ${\cal
T}=1$, all minigaps disappear.

\section{Kinetic equation} \label{sec-kinetic}

If a bias voltage $V$ is applied across the superconducting leads,
the current through the contact has both  ac and dc components.
For low voltages, $eV$ much smaller than $\Delta$, the dc current
is small for contacts with any transparency ${\cal T} \ne 1$. This
is due to the presence of minigaps discussed in the previous
Section. In long contacts the minigaps exist at $\phi = \pi k$ and
suppress the transitions between the levels thus preserving the
equilibrium distribution of excitations. As a result, the current
through the contact is simply the equilibrium supercurrent with
the phase difference $\phi$ adiabatically depending on time, $\phi
= \omega_J t+ \phi_0$ where $\omega_J=2eV/\hbar $ is the Josephson
frequency. The dc component should thus vanish. However, the time
dependence of the phase induces the Zener transitions between the
levels near the avoided crossing points at $\phi = \pi k$ which
produce deviation from equilibrium. Since the transparency of the
contact is of the order of unity, ${\cal T}\sim 1$, particles with
$|\epsilon|>|\Delta|$ have enough time to escape from the
double-barrier region and to relax in the continuum. Particles at
the continuum edges with energies $\epsilon =\pm |\Delta|$, which
are in equilibrium with the heat bath, are captured on the
outermost Andreev levels when the latter split off from the
continuum as the phase varies in time. Next these particles are
excited to the neighboring levels due to the interlevel
transitions. Relaxation of thus created nonequilibrium
distribution gives rise to a finite dc current.

In this Section we derive the effective kinetic equation that
describes the distribution function on the Andreev levels taking
into account both the inelastic relaxation on each level and the
Zener transitions between the neighboring levels. We assume that
the Zener transitions take place only near the avoided crossings
at $\phi = \pi k$. This can be realized if two conditions are
fulfilled. First, the minigaps $\delta \epsilon _\pi$ and $\delta
\epsilon _{2\pi}$ should be much smaller than the average distance
$\hbar v_x/d$ between the levels, implying that the contact is
almost ballistic with a transparency close to unity, $1-{\cal
T}\ll 1$. Second, the applied voltage should be small $eV \ll
\hbar v_x/d$ such that it cannot excite transitions between levels
far from the avoided crossings. We also assume that, between the
Zener tunneling events, the distribution function relaxes
according to
\[
\frac{\partial f_n}{\partial t}=-\frac{f_n-f^{(0)}_n}{\tau}
\]
where $\tau $ is an inelastic relaxation time which we assume
constant, and $f^{(0)}_n=1-2n_{\epsilon_n}^{(0)}=\tanh (\epsilon_n
/2T)$ is the equilibrium distribution. Here $n$ labels the levels
consecutively from the lowermost level at $\epsilon =-|\Delta|$ up
to the uppermost level at $\epsilon =+|\Delta|$. We will consider
low temperatures $T\ll \hbar v_x/d$ such that
\begin{equation}
f^{(0)}_n ={\rm sign}\, (\epsilon_n) \ .\label{equilib}
\end{equation}
Since the superconducting phase difference depends linearly on
time the distribution function can be written as
$f_n(\phi)=f_n^{(0)}+\tilde f_n$ where
\begin{equation}
\tilde f_n(\phi)=f_ne^{-\hbar \phi /2eV\tau} \ . \label{evolution}
\end{equation}
The amplitudes $f_n$ of the decaying parts are to be found using
the conditions at the transition points.
\begin{figure}[tbh]
\centerline{\includegraphics[width=0.5\linewidth]{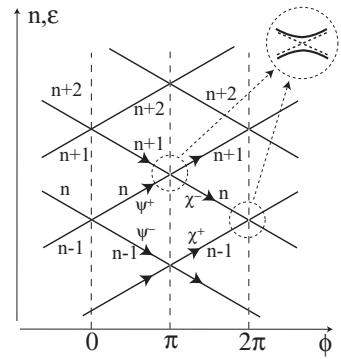}}
\caption{Scheme of the energy levels as functions of $\phi$. The
arrows show the direction of the spectral flow. Enlarged are shown
avoided crossings of levels at $\phi =2\pi k$ and $\phi =\pi +
2\pi k$. } \label{fig-boundcond}
\end{figure}

We assume that the interlevel transitions at $\phi =2\pi k$ have a
(constant) tunneling probability $p_0$ while the transitions at
$\phi =\pi +2\pi k$ have a probability $p_\pi$. We will see later
that the assumption of constant probabilities is well justified.
We denote the points $\phi =2\pi k$ before and after the tunneling
events as $0\mp$, respectively, while the points $\phi =\pi + 2\pi
k$ before and after the tunneling events are denoted as $\pi \mp$.
All the respective points $0\mp$ on a level $n$ are equivalent due
to the $2\pi$ periodicity, so are all the points $\pi \mp$. Using
the evolution equation (\ref{evolution}) we couple the
distribution functions at the consecutive instants of the
tunneling events (see Fig.~\ref{fig-boundcond})
\begin{eqnarray*}
f_n^{\pm}(\pi +)=f_n^{(0)}+\chi _n^{\pm} \ &,& \;
f_n^{\pm}(0-)=f_n^{(0)}+\chi _n^{\pm}e^{-\nu}\ , \\
f_n^{\pm}(0 +)=f_n^{(0)}+\psi _n^{\pm} \ &,& \; f_n^{\pm}(\pi
-)=f_n^{(0)}+\psi _n^{\pm}e^{-\nu}\ ,
\end{eqnarray*}
where $ \nu = \pi \hbar /2eV\tau $. Here we introduce the upper
$+$ (or $-$) indices to indicate explicitly the distributions at
the spectrum branches increasing (or decreasing) as functions of
$\phi$. The coefficients $\phi ^\pm _n$ and $\chi^\pm _n$ are the
amplitudes $f_n$ in Eq.\ (\ref{evolution}) defined for the
intervals $0<\phi<\pi$ and $\pi <\phi <2\pi$, respectively.

The tunneling events impose the relations
\begin{eqnarray}
f_n^+(0+)&=&p_0 f^+_{n-1}(0-)+(1-p_0)f_n^-(0-)\ , \label{f+0+}  \\
f_n^-(\pi +)&=&p_\pi f^-_{n+1}(\pi -)+(1-p_\pi )f_n^+(\pi -) \ ,
\label{f-pi+} \\
f_{n+1}^-(0+)&=&p_0 f^-_{n+2}(0-)+(1-p_0)f_{n+1}^+(0-) \ ,
\label{f-0+} \\
f_{n+1}^+(\pi +)&=&p_\pi f^+_{n}(\pi -)+(1-p_\pi)f_{n+1}^-(\pi -)\
, \label{f+pi+}
\end{eqnarray}
illustrated in Fig.~\ref{fig-boundcond}. For the coefficients
$\psi$ and $\chi$ they become
\begin{eqnarray}
\psi _n^{+}&=&p_0 [f_{n-1}^{(0)}-f_n^{(0)}]\nonumber \\
&&+[p_0\chi _{n-1}^{+}
+(1-p_0) \chi _n^{-}]e^{-\nu}  , \label{psi+n}\\
\psi _{n+1}^{-}&=&p_0[f_{n+2}^{(0)}-f_{n+1}^{(0)}]\nonumber \\
&&+[p_0\chi _{n+2}^{-} +(1-p_0)\chi _{n+1}^{+}]e^{-\nu}  ,
\label{psi-n}
\end{eqnarray}
and
\begin{eqnarray}
\chi _n^{-}=p_\pi [f_{n+1}^{(0)}-f_n^{(0)}]+[p_\pi \psi _{n+1}^{-}
+(1-p_\pi )\psi _n^{+}]e^{-\nu} ,\;\;
\label{chi-n} \\
\chi _{n+1}^{+}=p_\pi [f_{n}^{(0)}-f_{n+1}^{(0)}]+[p_\pi \psi
_{n}^{+} +(1-p_\pi)\psi _{n+1}^{-}]e^{-\nu} .\; \; \label{chi+n}
\end{eqnarray}

According to our picture of the spectral flow through the Andreev
levels, the boundary conditions are imposed at the continuum edges
in such a way that, for the levels increasing as functions of
$\phi$, the distribution $f^+$ coincides with the equilibrium at
$\epsilon =-|\Delta|$, while, for decreasing levels, the
distribution $f^- $ coincides with the equilibrium at $\epsilon
=+|\Delta|$. Since the bias voltage is low, $eV\ll \hbar v_x/d$,
the equilibrium function in both superconducting electrodes can be
taken as $f^{(0)}=\tanh(\epsilon_n/2T)$. Let us assume that
trapping of particles from the continuum occurs at $\phi =0$; the
boundary conditions are then formulated for the function $\psi^{+}
_n$ at $\epsilon =-|\Delta|$ and for $\psi^{-} _n$ at $\epsilon
=+|\Delta|$:
\begin{equation}
\psi^{+} _{\epsilon =-|\Delta|}=0 \, , \; \psi^{-} _{\epsilon
=+|\Delta|}=0 \ . \label{initialcond}
\end{equation}
In this case it is convenient to exclude the functions $\chi$
using Eqs.\ (\ref{chi-n}) and (\ref{chi+n}) and solve Eqs.\
(\ref{psi+n}), (\ref{psi-n}) for the functions $\psi$.

We choose the level index $n$ in such a way that $\epsilon_n >0$
for $n \geq 1$ and $\epsilon _n <0$ for $n \leq 0$. Equations
(\ref{psi+n}) and (\ref{psi-n}) then couple the levels $n$ and
$n\pm 2$. Since the temperature is low and the distribution is
given by Eq.\ (\ref{equilib}), the r.h.s. of these equations
vanish for all $n \ne 0,2$. Therefore, the coefficients for $n\geq
2$ and $n\leq 0$ satisfy the homogeneous equations. We assume that
there are $N+1$ levels with positive energies and $N+1$ levels
with negative energies such that the outermost levels touch the
continuum. Therefore, for the uppermost level $n=N+1$, the
solution of Eqs.\ (\ref{psi+n}), (\ref{psi-n}) satisfies the
condition $\psi ^-_{N+1}=0$. Similarly, for the lowermost level
$n=-N$ the solution satisfies $\psi ^+_{-N}=0$. With these
conditions, the solutions for $n\geq 2$ are
\begin{eqnarray}
\psi^+_n &=&c^>\left[ e^{-r(N-n+1)}w_- -e^{r(N-n+1)} w_+\right] \ ,
\nonumber \\
\psi^-_n &=&c^>\left[ e^{-r(N-n+1)} -e^{r(N-n+1)} \right] \ .
\label{psi>}
\end{eqnarray}
where
\[
w_\pm = \frac{p_0e^{\pm r}+p_\pi e^{\mp r}-2p_0p_\pi \cosh
(r)}{\zeta +p_0p_\pi (1-e^{\pm 2r})+p_0+p_\pi -2p_0p_\pi} \ .
\]
The solutions for $n\leq 0$ have the form
\begin{eqnarray}
\psi^+_n &=&c^<\left[ e^{r(N+n)}-e^{-r(N+n)}\right] \ , \nonumber
\\
\psi^-_n &=&c^<\left[ e^{r(N+n)}w_+ -e^{-r(N+n)}w_- \right] \ .
\label{psi<}
\end{eqnarray}

The effective relaxation rate $r>0$ is found from the determinant
condition $ w_+w_-=1 $ which gives
\begin{eqnarray}
4p_0p_\pi (\zeta +1)\sinh ^2(r)&=&\left(\zeta +p_0+p_\pi
-2p_0p_\pi\right)^2 \nonumber \\
&&- \left(p_0+p_\pi -2p_0p_\pi\right)^2 \ . \quad
\label{r-general}
\end{eqnarray}
where $\zeta =e^{2\nu} -1 $.

For an ideally transparent contact $p_0=p_\pi =1$ we find $r=\nu$.
For strong relaxation, $\nu \gg 1$, we also have $r\approx \nu$,
and the distribution relaxes quickly. The most interesting limit
for a general case $p_0,p_\pi \ne 1$ is when inelastic relaxation
is weak, $\nu \ll 1$. We find in this limit
\begin{equation}
\sinh^2 (r)=\nu (\nu +p_0+p_\pi- 2p_0p_\pi)/p_0p_\pi \ .
\label{r-expr}
\end{equation}
The relaxation rate $r$ can be either large or small depending on
the probabilities. The inverse rate $r^{-1}$ describes the
broadening of distribution over the energy states and plays the
role of an effective temperature $T_{\rm eff}=
r^{-1}(d\epsilon/dn)=\pi \hbar v_x/2rd$. The effective temperature
can be much higher than the interlevel spacing if $r\ll 1$.


The coefficients $c^>$ and $c^<$ are coupled through the solutions
of four non-homogeneous equations resulting from two equations
(\ref{psi+n}) and (\ref{psi-n}) taken for two values $n=2$ and
$n=0$. Inspecting equations for other $n$ we see that only the two
coefficients $\psi_0^+$ and $\psi_1^-$ cannot be described by the
solutions Eqs.\ (\ref{psi>}) and (\ref{psi<}) of the homogeneous
equations. We write
\begin{eqnarray}
\psi_1^-= \psi_1^{->} +\delta _1^> \ &,& \;
\psi_1^-= \psi_1^{-<} +\delta _1^< \ , \label{delta1}\\
\psi_0^+= \psi_0^{+>} +\delta _0^> \ &,& \;
\psi_0^+=\psi_0^{+<}+\delta _0^< \ . \label{delta0}
\end{eqnarray}
Here $\delta_{0,1}^>$ and $\delta_{0,1}^<$ are four new unknown
coefficients. The coefficients $\psi_{1}^{->}$ and $\psi_{0}^{+>}$
are defined to satisfy the homogeneous equations for $\epsilon
_n>0$ and are given by Eq.\ (\ref{psi>}); the coefficients
$\psi_1^{-<}$ and $\psi_0^{+<}$ satisfy the homogeneous equations
for $\epsilon _n<0$ and are given by Eq.\ (\ref{psi<}). Inserting
Eqs.\ (\ref{delta1}), (\ref{delta0}) into the four equations
obtained for $n=2$ and $n=0$ from Eqs.\ (\ref{psi+n}) and
(\ref{psi-n})  we find all the four coefficients $\delta$. The
result is $\psi_1^-= \psi_1^{->}$, $\psi_0^+=\psi_0^{+<}$ while
\begin{eqnarray}
\psi_1^{-<}-\psi_1^{->}=e^{\nu}( f^{(0)}_{1}-f^{(0)}_{0})\ , \label{jump1}\\
\psi_0^{+<}-\psi_0^{+>}=e^{\nu}( f^{(0)}_{1}-f^{(0)}_{0})\ .
\label{jump2}
\end{eqnarray}
These two equations yield $c^>=c^<=C$ where
\begin{equation}
C\left[ e^{rN}\left( 1 +e^{r}w_+ \right) -e^{-rN}\left(1+e^{-r}w_-
\right)\right] =2e^\nu \ . \label{C-coeff}
\end{equation}
We put $f^{(0)}_{1}-f^{(0)}_{0}=2$ for low temperatures.
Therefore, the distribution possesses the symmetry $\psi
_{-n}^+=-\psi _{n+1}^-$.

\section{Current} \label{sec-current}

The contribution to the current due to the deviation from
equilibrium is
\begin{equation}
I_{\rm neq}=-\frac{2e}{\hbar}\sum _{\epsilon _n >0} \frac{\partial
\epsilon _n}{\partial \phi} \tilde f_n
\end{equation}
where $\tilde f_n = f_n -f^{(0)}_n$. The sum runs only over the
localized Andreev states because the continuum states relax
quickly so that their distribution is almost in equilibrium. The
equilibrium supercurrent has been calculated in
Ref.~[\onlinecite{GalZaik02}] (see also
Ref.~[\onlinecite{Kupriyanov}] for a review); it is an oscillating
function of the phase difference and thus has no contribution to
the dc current.

Denote $ \partial \epsilon _n^\pm /\partial \phi$ increasing
(decreasing) parts of the spectrum $\epsilon_n (\phi)$ as a
function of $\phi$. We have for the current averaged in time
\begin{eqnarray}
\overline{I} &=& -\frac{e}{\pi \hbar} \sum _{l=0}^{N/2} \left[
\int_0^\pi \left(\frac{\partial \epsilon_{n+1} ^+}{\partial \phi}
\tilde f^+_{n+1}+ \frac{\partial \epsilon_n ^-}{\partial \phi}
\tilde f^-_{n}\right)\, d\phi
\right. \nonumber \\
&& + \left.\int_{\pi}^{2\pi}  \left(\frac{\partial
\epsilon_{n+1}^-}{\partial \phi}\tilde f^-_{n+1}+\frac{\partial
\epsilon_n ^+}{\partial \phi}  \tilde f^+_{n}\right)\, d\phi
\right]_{n=2l+1}  . \quad \label{current1}
\end{eqnarray}
The sum over $l$ runs from 0 to $L=N/2$ where
\[
N=2d\Delta /\pi \hbar v_x
\]
is the total number of levels with $\epsilon_n>0$, i.e., for both
signs in Eq.\ (\ref{spectrum1}) .

In Eq.\ (\ref{current1}) we can use the ballistic spectrum with
${\cal T}\rightarrow 1$. In this limit $|S_N|=0$, $|S_A|=1$, thus
the spectrum in Eq.\ (\ref{determinant2}) takes the form
\[
\cos (\phi /2) = \pm \sin (\beta +\gamma) \ .
\]
Calculating the energy derivative of this equation for long
junctions $d|\Delta| \gg \hbar v_x$, we find
\[
\frac{\partial \epsilon_n ^\pm }{\partial \phi}=\pm \frac{\hbar
v_x}{2d} \ .
\]
We neglected $\partial \gamma /\partial \epsilon$ compared to
$\partial \beta /\partial \epsilon$ which holds for all energy
levels excluding those in a narrow region near the gap edge,
$1-|\epsilon | /|\Delta |\ll (\hbar v_x/d|\Delta|)^2$. This means
in fact that, neglecting this  narrow region, we can use  Eqs.\
(\ref{spectrum1}) and (\ref{spectrum-sym}) with ${\cal T}=1$ for
all $n$.

We have for $0<\phi <\pi$
\[
\tilde f_n ^\pm =\psi ^\pm _n e^{-\hbar \phi /2eV\tau} \ ,
\]
while for $\pi <\phi <2\pi$
\[
\tilde f_n ^\pm =\chi ^\pm _n e^{-\hbar (\phi -\pi) /2eV\tau} \ .
\]
We obtain for $\nu \ll 1$
\begin{equation}
\overline{I} =\frac{ev_x}{2d}\Phi(p_0,p_\pi)=\frac{\pi \hbar
v_x}{2eR_0d}\Phi(p_0,p_\pi) \label{current2}
\end{equation}
where $R_0^{-1}=e^2/\pi \hbar$ is the quantum of conductance, and
\[
\Phi(p_0,p_\pi) =-\sum _{l=0}^{N/2} \left[ \left(\psi^+_{n+1}-
\psi^-_{n}\right)+ \left(\chi^+_{n}-
\chi^-_{n+1}\right)\right]_{n=2l+1} . 
\]


The combination of coefficients $\psi$ and $\chi$ that enters the
expression for the supercurrent for $l\geq 1$ can be written
through the solutions Eq.\ (\ref{psi>}) of the homogeneous
equations (\ref{psi+n})--(\ref{chi+n}). The term $l=0$ contains
$\chi ^+_{1}$ which is expressed through $\psi ^+_0$. However, one
can check that the jump in $\psi _n^+$ from $n=0$ to $n=2$ is
compensated by the jump $f_1^{(0)}-f_0^{(0)}$ in Eq.\
(\ref{chi+n}).

Consider the limit of low relaxation $r\ll 1$ provided $Nr\gg 1$.
The limit $r\ll 1$ is realized when $\nu \ll p_0,p_\pi$. In this
case Eqs.\ (\ref{chi+n}), (\ref{psi>}), and (\ref{C-coeff}) give
\begin{equation}
(\psi ^+_{n+1} -\psi ^-_{n})+(\chi^+_{n}-\chi^-_{n+1})=-\frac{4\nu
}{r}e^{-rn}\ . \label{psi-comb1}
\end{equation}
We have from Eq.\ (\ref{psi-comb1})
\begin{equation}
\Phi(p_0,p_\pi)=\frac{4\nu}{r}\sum _{k=0}^{N/2}e^{ -r(2k+1)}
=\frac{2p_0p_\pi}{p_0+p_\pi -2p_0p_\pi +\nu} \ . \label{Phi1}
\end{equation}
We keep $\nu$ in the denominator since the combination
$p_0+p_\pi-2p_0p_\pi$ vanishes when $p_{0},p_{\pi}\rightarrow 1$.

When the inelastic relaxation rate is so small that $N\nu \ll 1$
the effective relaxation $r$ can decrease such that $Nr \ll 1$.
Since the product $N(p_0+p_\pi -2p_0p_\pi )$ is generally not
small we find from Eqs.\ (\ref{chi-n}), (\ref{chi+n}),
(\ref{psi>}), and (\ref{C-coeff})
\[
(\psi ^+_{n+1} -\psi
^-_{n})+(\chi^+_{n}-\chi^-_{n+1})=-\frac{4p_0p_\pi}{N(p_0+p_\pi
-2p_0p_\pi )+1},
\]
and
\begin{equation}
\Phi(p_0,p_\pi)=\frac{2p_0p_\pi}{(p_0+p_\pi -2p_0p_\pi )+1/N}\ .
\label{Phi2}
\end{equation}
Eqs.\ (\ref{Phi1}) and (\ref{Phi2}) go one into another for $N\sim
1/\nu $. The exact expression for $\Phi$ is found from Eqs.\
(\ref{chi-n}), (\ref{chi+n}), (\ref{psi>}), and (\ref{C-coeff}).
One can approximate the function $\Phi(p_0,p_\pi)$ by an
interpolation between Eqs.\ (\ref{Phi1}) and (\ref{Phi2}) in the
form
\begin{equation}
\Phi(p_0,p_\pi)=\frac{2p_0p_\pi}{(p_0+p_\pi -2p_0p_\pi )+2\beta}
\label{Phi-final}
\end{equation}
where $ 2\beta \approx {\rm max}\, (\nu , N^{-1}) \approx \nu +
N^{-1} $.

The probability of Zener tunneling can be easily calculated for
the spectrum in the form of Eqs.\ (\ref{spectrum1}),
(\ref{spectrum-sym}) if $(1-{\cal T}^2)\ll 1$. The phase
difference is $ \phi = \omega _Jt + \phi_0 $. As a function of
time, the distance between two neighboring levels for $\phi $
close to $\pi$ is
\[
\delta \epsilon =\frac{\hbar v_x{\cal T}\omega _J}{d}\sqrt{
t^2+\tau_0^2}
\]
where $t$ is small and
\[
\tau_0^2=4\sin ^2\alpha ^\prime(1-{\cal T}^2)/{\cal T}^2\omega
_J^2 \ .
\]
Probability of Zener tunneling is
\[
p_\pi =\exp \left[ -\frac{2}{\hbar}\, {\rm Im} \left(\int
_0^{i\tau_0}\delta \epsilon \, dt \right)\right]= \exp \left[
-\frac{\omega _0}{\omega _J}\sin^2\alpha^\prime \right]
\]
where
\[
\omega_0=\pi v_x(1-{\cal T}^2)/{\cal T}d \ .
\]
The distance between two levels for $\phi$ close to $\phi =0$ and
the corresponding probability of Zener tunneling $p_0$ are given
by the same expressions with $\sin \alpha^\prime$ replaced with
$\cos \alpha^\prime$.

In Eq.\ (\ref{Phi-final}) the term with $\beta$ is only important
when $p_0$ and $p_\pi $ are close to unity. Therefore, one can
write
\begin{equation}
\Phi (p_0,p_\pi)
=\left(\exp\left[\frac{\omega_0}{\omega_J}\right]\cosh
\left[\frac{\omega _0}{\omega _J}\cos (2\alpha^\prime)\right]
-1+\beta\right)^{-1} . \label{interpol}
\end{equation}

When the bias voltage is low $\omega_J/\omega _0 \lesssim 1$, such
that $\nu \ll p_0,p_\pi \ll 1$, we have
\begin{equation}
\Phi (p_0,p_\pi) =2\exp\left(-\frac{\omega_0}{\omega_J}[1+|\cos
(2\alpha^\prime)|]\right) \ .
\end{equation}
The low-voltage part is exponential due to small Zener
probabilities. The exponent exhibits strong oscillations as a
function of $\alpha^\prime$ which can be manipulated by varying
the gate voltage.

For higher bias voltages, $\omega_J/\omega _0 \gg 1$, we have
$p_{0},p_{\pi}\rightarrow 1$ and
\[
\Phi (p_0,p_\pi) = \frac{\omega
_J}{\omega _0 +\omega_J \beta}\approx \frac{\omega _J}{\omega
_0+\pi/2\tau +\omega_J/2N} \ .
\]
For a fully ballistic contact with $p_0=p_\pi =1$, i.e.,
$\omega_0=0$ our result agrees qualitatively with
Ref.~[\onlinecite{GunsZaik94}]. For these voltages we have two
regimes. The I-V curve is linear $I=V/R$ as long as $eV\ll N(\hbar
\omega _0+\pi \hbar /\tau )$. The effective conductance is
\[
\frac{1}{R}=\frac{1}{R_0}\frac{\pi v_x/d}{\omega _0+\pi/2\tau}\ .
\]
Inelastic relaxation can be neglected if $ \omega_0\gg \pi /2\tau
$. In this case the conductance is much larger than the
conductance quantum $R_0/R= (1-{\cal T}^2)^{-1}$. It is
interesting to note that the effective conductance is independent
of the gate voltage: it contains the sum of two functions in the
exponents for $p_0$ and $p_\pi$, i.e.,
$(\omega_0/\omega_J)\cos^2\alpha^\prime $ and
$(\omega_0/\omega_J)\sin^2\alpha^\prime $, which obviously is
independent of $\alpha^\prime$.

With increasing voltage up to $eV \gtrsim N(\hbar \omega _0+\pi
\hbar /\tau )$ the I-V curve saturates at the value
\[
I=Nev_x/d =2e|\Delta|/\pi \hbar
\]
which is by a factor $N\gg 1 $ larger than the critical Josephson
current of the junction \cite{GalZaik02}, $I_c \sim ev_x/d$.

\section{Conclusions} \label{sec-discuss}

To summarize, we have considered the low temperature charge
transport in a nearly ballistic single-mode SINIS junction having
a length $d$ longer than the superconducting coherence length
$\xi$. In this junction the energy spectrum of Andreev states has
a large number of levels separated from each other by minigaps
which do not vanish in a realistic case when the transmission is
not exactly unity. In the limit of low bias voltages, we have
derived and solved the kinetic equation for the quasiparticle
distribution on the Andreev levels that takes into account both
inelastic relaxation and voltage-driven Zener transitions between
the levels. We have shown that the Zener transitions enhance the
action of each other and lead to a drastic increase of the dc
current. The voltage dependence of the dc current is first
exponential due to small probabilities of Zener tunneling. Next it
goes over into a linear relation such that, at low temperatures
when the inelastic relaxation rate is slow, its slope is
determined by the average minigap in the spectrum. At higher
voltages when the Zener probabilities approach unity, the dc
current saturates at a value far exceeding the critical Josephson
current of the junction.

\acknowledgments

We are grateful to Yu. Galperin and A. Mel'nikov for stimulating
discussions. The work was supported in part by the Russian
Foundation for Basic Research (grant 06-02-16002-a).


\begin{thebibliography}{99}

\bibitem{Kupriyanov} A.A.\ Golubov, M.Yu.\ Kupriyanov and E.\
Il'ichev, Rev.\ Mod.\ Phys.\ {\bf 76}, 411 (2004).

\bibitem{Andreev} A.F. Andreev, Zh. Eksp. Teor. Fiz. {\bf 46}, 1823 (1964)
[Sov. Phys. JETP {\bf 49}, 924 (1964)]


\bibitem{GunsZaik94} U. Gunsenheimer and A.D. Zaikin, Phys. Rev. B
{\bf 50}, 6317 (1994).

\bibitem{AverBard95} D. Averin and A. Bardas, Phys. Rev. Lett. {\bf 75}, 1831
(1995).

\bibitem{AverBard96} D. Averin and A. Bardas, Phys. Rev. B {\bf 53}, R1705
(1996).
\bibitem{BTK} G.E.\ Blonder, M.\ Tinkham, and T.M.\ Klapwijk, Phys.\
Rev.\ B, {\bf 25}, 4515 (1982).

\bibitem{Octavio83} M. Octavio, M. Tinkham, G.E. Blonder, and T.M.
Klapwijk, Phys. Rev. B {\bf 27}, 6739 (1983).

\bibitem{KV03} N. B. Kopnin and V. M. Vinokur, Europhys. Lett. {\bf 61},
824 (2003).

\bibitem{nanotubes} M.R.\ Buitelaar, W.\
Belzig, T.\ Nussbaumer, et al.,  Phys.\ Rev.\ Lett.\ {\bf 91},
057005 (2003); A.\ Kosumov, M.\ Kociak, M.\ Ferrier, et al.,
Phys.\ Rev.\ B {\bf 68}, 214521 (2003); P.\ Jarillo-Herrero, J.A.\
van Dam, and L.P.\ Kouwenhoven, Nature {\bf 439}, 953 (2006).

\bibitem{nanowire}  Y.J.\ Doh, J.A.\ van Dam , A.L.\ Roest, et
al., Science {\bf 309}, 272 (2005).

\bibitem{KMV06} N.B. Kopnin, A.S. Mel'nikov, and V.M. Vinokur,
Phys. Rev. Lett. {\bf 96}, 146802, (2006).

\bibitem{Beenakker} C.W.J.\ Beenakker, Phys.\ Rev.\ Lett.\ {\bf 67},
3836, (1991).

\bibitem{spectrum} U.\ Sch{\"u}ssler and R.\ K{\"u}mmel, Phys.\ Rev.\ B {\bf 47},
2754 (1993); G.A.\ Gogadze and A.M.\ Kosevich, Fiz.\ Nizkih Temp.\
{\bf 24}, 716 (1998) [Low Temp.\ Phys.\ {\bf 24}, 540 (1998)]; A.\
Jacobs and R.\ K{\"u}mmel,  Phys.\ Rev.\ B {\bf 71}, 184504
(2005); D.D.\ Kuhn, N.M.\ Chtchelkatchev, G.B.\ Lesovik, and G.\
Blatter, Phys.\ Rev.\ B, {\bf 63}, 054520 (2001).

\bibitem{GalZaik02} A.V.\ Galaktionov and A.D.\ Zaikin, Phys.\ Rev.\ B
{\bf 65}, 184507 (2002).

\end{thebibliography}
\end{document}